\documentclass[prd,superscriptaddress,twocolumn,nofootinbib,showpacs]{revtex4-1}
\usepackage{amsfonts,amssymb,amsmath}
\usepackage{mathrsfs}
\usepackage{color}
\usepackage{graphicx}
\usepackage{dcolumn}
\usepackage{bm}

\usepackage{pslatex}
\usepackage[T1]{fontenc}
\usepackage[latin1]{inputenc}
\usepackage{graphicx}
\usepackage{epsfig}
\usepackage{longtable}
\usepackage{float}
\usepackage{calc}
\usepackage{ifthen}

{
{
{
\newcommand{\bea}{\begin{eqnarray}}
\newcommand{\eea}{\end{eqnarray}}

\newcommand{\nc}{\newcommand}
\nc{\renc}{\renewcommand}
\nc{\eqs}[2]{\mbox{Eqs.~(\ref{#1},\,\ref{#2})}}
\nc{\eq}[1]{\mbox{Eq.~(\ref{#1})}}
\nc{\figs}[2]{\mbox{Figs.~(\ref{#1},\,\ref{#2})}}
\nc{\fig}[1]{\mbox{Fig~.(\ref{#1})}}
\nc{\be}[1]{\begin{equation} \mbox{$\label{#1}$}}
\nc{\ee}{\vspace{0.1cm}\end{equation}}

\newcommand{\bean}{\begin{eqnarray*}}
\newcommand{\eean}{\end{eqnarray*}}

\def\bfx{{\bf x}}

\begin{document}
\title{Inflaton condensate fragmentation: Analytical conditions and application to $\alpha$-attractor models}
\author{Jinsu Kim}
\email{kimjinsu@kias.re.kr}
\affiliation{Quantum Universe Center, Korea Institute for Advanced Study, Seoul 02455, Korea}
\author{John McDonald}
\email{j.mcdonald@lancaster.ac.uk}
\affiliation{Dept. of Physics, Lancaster University, Lancaster LA1 4YB, UK}

\begin{abstract}

We study the stability of an inflaton condensate in the presence of attractive inflaton self-interactions, in order to determine analytical conditions on the self-interaction couplings under which the condensate undergoes fragmentation. As an application of our results, we consider the stability of the inflaton condensate in $\alpha$-attractor inflation. We show that the stability of the condensate depends upon the value of $\alpha$. For the T-model with $n = 1$, we find that it is unstable for $\alpha \lesssim 10^{-4}$. In this case it is expected that inflation will be followed by an oscillon-dominated era.

\end{abstract}

\maketitle

\section{Introduction}
\label{sec:intro}

Inflation typically ends with the formation of a condensate described by a coherently oscillating scalar field. In the presence of an attractive inflaton self-interaction, the condensate can fragment into lumps, which are expected to evolve into oscillons \cite{jm1,iballs,amin}. (For very rapid perturbation growth, fragmentation can also occur via tachyonic preheating \cite{tp,cope,jm2}.)  Our aim in this paper is to establish conservative analytic bounds on the self-couplings under which fragmentation of an inflaton condensate will definitely occur. To do this, we will assume the existence of a coherently oscillating condensate and then determine the strength of self-interaction which causes the growth of primordial inflaton perturbations to fragment the condensate. We first present a general method to analyze the growth of perturbation modes in the case where the rate of growth is slow compared to the frequency of oscillation and the potential is dominated by a quadratic term. This is closely related to the method of Ref. \cite{lee} (see also Ref. \cite{kus}). We then apply our method to the case of cubic and quartic self-interactions to determine lower bounds on their couplings for the condensate to fragment. 
As a specific application of our results, we consider the case of T-model $\alpha$-attractor inflation  \cite{Kallosh:2013,Galante:2014ifa,Kallosh:2015}. We will show that the condensate in this model will fragment for sufficiently small values of $\alpha$.

\section{Perturbation Growth in Real Scalar Condensates}
\label{sec:pert}
A real scalar field, denoted by $\Phi$, is governed by the equation of motion,
\begin{align}
	\ddot{\Phi} + 3H\dot{\Phi} - \frac{\nabla^{2}\Phi}{a^{2}} + \frac{\partial V}{\partial \Phi} = 0
	\,,
\end{align}
where $H\equiv \dot{a}/a$ is the Hubble rate, $a$ is the scale factor and $V$ is the scalar potential.
Defining $\Phi = (a_{0}/a)^{3/2}\phi$, where $a=a_{0}$ is a reference scale factor at which $\Phi = \phi$, we obtain
\begin{align}\label{eqn:eomphi}
	\ddot{\phi} - \frac{\nabla^{2}\phi}{a^{2}}
	=
	-\frac{\partial U(\phi)}{\partial \phi}
	\,.
\end{align}
Here
\begin{align}
	\frac{\partial U}{\partial \phi} \equiv
	\left(
		\frac{a}{a_{0}}
	\right)^{3/2}
	\frac{\partial V}{\partial \Phi}
	+\Delta_{H}\phi
	\,,
\end{align}
with
\begin{align}
	\Delta_{H} \equiv
	-\frac{3}{2}\left(
		\dot{H} + \frac{3}{2}H^{2}
	\right)\,.
\end{align}
We assume that $\phi$ is periodic and write $\phi({\bf x}, t) = R \sin\Omega$, where $R$ and $\Omega$ are functions of ${\bf x}$ and $t$. The $\sin \Omega$ term determines the periodic nature of the $\Phi$ field and allows for space-dependent fluctuations of the frequency. 
In the absence of expansion and perturbations, $R$ is a constant and $\Omega = \Omega(t)$.  Therefore our analysis applies to potentials which are symmetric under $\Phi \leftrightarrow -\Phi$. We then perturb about this background field.

Substituting into Eq. \eqref{eqn:eomphi} gives
\begin{align}
	&\left(
		\ddot{R} - R\dot{\Omega}^{2}
		-(\nabla^{2}R)/a^{2}
		+R(\nabla\Omega)^{2}/a^{2}
	\right)\sin\Omega
	\nonumber\\
	&
	+\left(
		2\dot{R}\dot{\Omega}
		+R\ddot{\Omega}
		-2(\nabla R \cdot \nabla \Omega)/a^{2}
		-R(\nabla^{2}\Omega)/a^{2}
	\right)\cos\Omega
	\nonumber\\
	&=
	-\frac{\partial U}{\partial\phi}
	\,.
\end{align}
We will restrict our analysis to the case where the potential is dominated by a quadratic term, so that the background oscillations are nearly harmonic. In this case, on the time scale of the oscillations, we can consider $\dot{\Omega}$, $R$ and their derivatives to be approximately constant. We will also consider the expansion rate to be small compared to the frequency of oscillation, so that $R$ does not change significantly due to expansion during an oscillation cycle. Finally, we will assume that $\nabla \Omega$ is independent of $\Omega$, which will be justified by the resulting solution for $\Omega(\bfx,t)$.

Multiplying by $\sin\Omega$ and averaging over oscillations, we obtain
\begin{align}\label{eqn:perteom1}
	&\ddot{R} - R\dot{\Omega}^{2}
	-(\nabla^{2}R)/a^{2}
	+R(\nabla\Omega)^{2}/a^{2}
	=
	-U_{{\rm eff}}^{\prime}
	\,,
\end{align}
where
\begin{align}\label{eqn:effPot}
	U_{{\rm eff}}^{\prime} \equiv
	\frac{\partial U_{{\rm eff}}}{\partial R}
	\equiv
	\frac{2}{R}\bigg\langle
		\phi\frac{\partial U}{\partial\phi}
	\bigg\rangle\,,
\end{align}
and
\begin{align}
	\langle f(\Omega) \rangle
	\equiv \frac{1}{2\pi}\int_{0}^{2\pi}
	d\Omega \, f(\Omega)
	\,.
\end{align}
Similarly, multiplying by $\cos\Omega$ and averaging over coherent oscillations, we obtain
\begin{align}\label{eqn:perteom2}
	\ddot{\Omega}
	+2\dot{R}\dot{\Omega}/R
	-2\nabla R \cdot \nabla\Omega / (a^{2}R)
	-(\nabla^{2}\Omega)/a^{2} = 0\,.
\end{align}

We next perturb $R$ and $\Omega$ about their homogeneous background values $R_{0}(t)$ and $\Omega_{0}(t)$,
\begin{align}\label{eqn:ROmPertParts}
	R(\bfx, t) &= R_{0}(t) + \delta R(\bfx, t)
	\,,\nonumber\\
	\Omega(\bfx, t) &= \Omega_{0}(t) + \delta \Omega(\bfx, t)
	\,.
\end{align}
$\delta R$ and $\delta \Omega$ then satisfy the following perturbation equations:
\begin{align}
	\delta \ddot{R}
	-2R_{0}\dot{\Omega}_{0}\delta \dot{\Omega}
	-\dot{\Omega}_{0}^{2}\delta R
	-\frac{1}{a^{2}}\nabla^{2}\delta R
	+U_{{\rm eff}}^{\prime\prime}	\delta R & =0
	\,,\label{eqn:eqnfordR}
	\\
	\delta \ddot{\Omega}
	-\frac{1}{a^{2}}\nabla^{2}\delta\Omega
	-\frac{2\dot{R}_{0}\dot{\Omega}_{0}}{R_{0}^{2}}\delta R
	+\frac{2\dot{R}_{0}}{R_{0}}\delta \dot{\Omega}
	+\frac{2\dot{\Omega}_{0}}{R_{0}}\delta \dot{R} & =0
	\,.\label{eqn:eqnfordOm}
\end{align}
Following Ref. \cite{lee}, we look for a solution to Eqs. \eqref{eqn:eqnfordR} and \eqref{eqn:eqnfordOm} of the form\footnote{The analysis presented in Ref. \cite{lee} considered perturbations of a globally charged complex scalar field rather than the real oscillating scalar field we are considering here.} 
\begin{align} \label{eqn:pertdRdOm}
	\delta R = \delta R_{0} e^{S(t) - i\mathbf{k}\cdot \mathbf{x}} \,,\quad \delta \Omega
	= \delta \Omega_{0} e^{S(t) - i\mathbf{k}\cdot \mathbf{x}}
	\,.
\end{align}
We will refer to $S(t)$ as the growth factor in the following.\footnote{The perturbation mode with wave number $k$ can be considered to represent a perturbation of the field characterized by a range of modes $\Delta k \sim k$ centered on $k$, with characteristic physical length scale $\delta x \sim \lambda_{{\rm phys}} = 2 \pi a/k$.} 
Since we are neglecting expansion on the time scale of the oscillations, we can set $H = \dot{H} = \dot{R}_{0} = 0$ for now. We will then look for a solution such that the rate of growth of the perturbations, $\alpha = \dot{S}$, is constant on time scales short compared to the expansion rate of the Universe. 
On substituting $\delta R$ and $\delta \Omega$ from Eq. \eqref{eqn:pertdRdOm} into Eqs. \eqref{eqn:eqnfordR} and \eqref{eqn:eqnfordOm} we obtain the dispersion relation
\begin{align} \label{eqn:dispersion}
	\left[
		\alpha^{2}
		+\frac{k^{2}}{a^{2}}
		+U_{{\rm eff}}^{\prime\prime} - \dot{\Omega}_{0}^{2}
	\right]\left[
		\alpha^{2}
		+\frac{k^{2}}{a^{2}}
	\right]
	+4\alpha^{2}\dot{\Omega}_{0}^{2} = 0
	\,.
\end{align}
This is similar to the dispersion relation in Ref. \cite{lee}. 
A solution for $\alpha$ which is real is only possible if the first term can become negative. This requires that 
\begin{align}\label{eqn:kc}
	\frac{k^2}{a^{2}} < 
	\dot{\Omega}_{0}^{2}
	-
	U_{{\rm eff}}^{\prime\prime}
	\equiv \frac{k_{c}^2}{a^{2}}\,.
\end{align}
In this we have defined $k_{c}$ to be the upper bound on the comoving wave number for growth of perturbations to be possible.

Solving Eq. \eqref{eqn:dispersion}, we obtain, for small $k/a$ values,
\begin{align}\label{eqn:alphasoln}
	\alpha^{2} &\approx
	\frac{k^{2}}{a^{2}}
	\frac{\dot{\Omega}_{0}^{2}-U_{{\rm eff}}^{\prime\prime}}{3\dot{\Omega}_{0}^{2}+U_{{\rm eff}}^{\prime\prime}} -16\frac{k^{4}}{a^{4}}\frac{\dot{\Omega}_{0}^{4}}{(3\dot{\Omega}_{0}^{2}+U_{{\rm eff}}^{\prime\prime})^{3}}
	\,.
\end{align}
We have assumed that the oscillations are dominated by the quadratic term in the potential. In this case we expect $\dot{\Omega}^{2}_{0} \approx U^{\prime\prime}_{{\rm eff}}$. In addition, we can use the equation for the background field $\Omega_{0}(t)$ from Eq. \eqref{eqn:perteom2}, which gives $\dot{\Omega}_{0}^{2} = U_{{\rm eff}}^{\prime}/R$ in the absence of expansion. In this case Eq. \eqref{eqn:alphasoln} becomes
\begin{align}\label{eqn:alphakc}
	\alpha^{2} \approx
	\frac{k^{2}}{a^{2}} \left(1 - \frac{k^2}{k_{c}^2}\right)
	\frac{U_{{\rm eff}}^{\prime}/R-U_{{\rm eff}}^{\prime\prime}}
	{4 U_{{\rm eff}}^{\prime\prime}}
	\,.
\end{align}
From this we obtain $\alpha(k, t)$ for a given mode $k$ at a given time $t$ on a time scale which is short compared to the expansion rate. We can then integrate the resulting $\alpha(k,t)$ over a period longer than the expansion time $H^{-1}$ to obtain the total growth factor in the expanding Universe for a given mode $k$,
\begin{align}\label{eqn:growthfactor}
	S(k, a(t)) = \int_{t_{0}}^{t} dt\,\alpha(k,t)
	= \int_{a_{0}}^{a(t)}\,da\,\frac{\alpha(k,a)}{aH}
	\,.
\end{align}

In order for fragmentation to occur, we require that the perturbation grows to nonlinearity, $\delta R / R \approx (\delta R_{0}/R_{0})\exp(S) \sim 1$, corresponding to  $S \sim \ln(R_{0}/\delta R_{0})$. The initial perturbation of the scalar field will be due to the primordial perturbation of the inflaton field, and therefore $\delta R_{0}/R_{0} \sim 10^{-4}$. Therefore the condition for the fragmentation to occur is given by
\begin{align}\label{eqn:fragcond}
	S(k, a(t)) > 10	\,.
\end{align}

This gives a general method for calculating the growth of perturbations of a condensate of scalars which have an attractive self-interaction. We next apply this method to the leading-order cubic and quartic interactions which can be expected to arise in the polynomial expansion of a general potential.

\section{Condensate Instability due to Cubic and Quartic Attractive Self-Interactions}
\label{sec:general}

\subsection{Cubic self-interaction}
\label{subsec:cubic}
We consider the potential
\begin{align}\label{eqn:CubicPot}
	V(\Phi) = \frac{1}{2} m^2 |\Phi|^2 - A |\Phi|^{3} \,,
\end{align}
where $A > 0$. The corresponding effective potential from Eq. \eqref{eqn:effPot} is 
\begin{align}
	U_{{\rm eff}}(R)
	=
	\frac{1}{2}m^{2}R^{2}
	-\frac{8A}{3\pi}
	\left(
		\frac{a_{0}}{a}
	\right)^{3/2}R^{3}
	\,,
\end{align}
where we have used $\langle |\sin^{3} \Omega|\rangle = 4/3\pi$. 

$k_{c}$ is given by
\begin{align}\label{eqn:maxkcubic}
	k_{c} &\approx
	\sqrt{\frac{8AR}{\pi}}
	a_{0}^{3/4}a^{1/4}\,.
\end{align}
The maximum comoving wave number for growth satisfies $k_{c} \propto a^{1/4}$. We are considering the growth of a mode with a fixed value of the comoving wave number $k$. Therefore, as $a$ increases, eventually a wave number that initially satisfies $k > k_{c}$ will become smaller than $k_{c}$, at which time the growth of the mode will begin. Thus, for a given $a_{0}$, the mode which subsequently has the greatest growth will correspond to $k = k_{c}(a_{0}) \equiv k_{c\,0}$. 
The growth factor is then 
\begin{align}
	S(k_{c\,0}, a) = \sqrt{\frac{2 A R}{\pi m^2}}
	\frac{k_{c\,0}}{a_{0}H_{0}} \int_{1}^{x} 
	x^{-5/4} \sqrt{1 - x^{-1/2}} \, dx
	\,,
\end{align}
where $x = a/a_{0}$.
In the limit $x \rightarrow \infty$, the integral is equal to $\pi$. Therefore the maximum growth factor is 
\begin{align}
	S_{{\rm max}} = \sqrt{\frac{2 A R}{\pi m^2}}
	\frac{\pi}{H_{0}}  \frac{k_{c\,0}}{a_{0}} = \frac{4 AR}{m H_{0}}
	\,.
\end{align}
The potential is assumed to be dominated by the quadratic term, and therefore
\begin{align}
	H_{0} = \sqrt{\frac{m^2 R^{2}/2}{3 M_{P}^2}} = \frac{m R}{\sqrt{6} M_{P}}
	\,.
\end{align}
Thus 
\begin{align}\label{eqn:SmaxCubic}
	S_{{\rm max}} = \frac{4 \sqrt{6} A M_{P}}{m^2}\,.
\end{align}
The condition for fragmentation, $S_{{\rm max}} > 10$, is therefore satisfied if 
\begin{align}
	A > \frac{5}{2 \sqrt{6}} \frac{m^2}{M_{P}}
	\,.
\end{align}

This is a conservative condition for fragmentation, since we have restricted to modes which begin growth during inflaton oscillations dominated by a quadratic potential. Therefore if the coupling $A$ satisfies this condition then fragmentation of the inflaton condensate will definitely occur.

\subsection{Quartic self-interaction}
\label{subsec:quartic}
We consider the potential
\begin{align}\label{eqn:QuarticPot}
	V(\Phi) = \frac{1}{2} m^2 \Phi^2 - A \Phi^{4} \,,
\end{align}
where $A > 0$. The corresponding effective potential is 
\begin{align}
	U_{{\rm eff}}(R)
	=
	\frac{1}{2}m^{2}R^{2}
	-\frac{3}{4}A\left(
		\frac{a_{0}}{a}
	\right)^{3}R^{4}
	\,,
\end{align}
where we have used $\langle \sin^{4} \Omega \rangle = 3/8$. 

$k_{c}$ is given by
\begin{align}\label{eqn:maxkquartic}
	k_{c} &\approx
	\sqrt{6A}R
	a_{0}^{3/2}a^{-1/2}\,.
\end{align}
In this case $k_{c} \propto a^{-1/2}$. As a result, a mode of a given wave number $k$ grows in a quite different way from the cubic case. Here, if $k$ is initially less than $k_{c}$ and so the mode is initially growing, then as $a$ increases and so $k_{c}$ decreases, $k$ will eventually become larger than $k_{c}$ and the growth of the mode will stop. 

In order to determine the maximum possible growth of a mode in this case, we first consider an initial scale factor $a_{0}$ at which $k < k_{c\,0}$. We then compute the growth factor $S(k,a_{c})$ at a later scale factor $a_{c}$ at which $k = k_{c}$ and the growth of the mode stops. We then need to maximize $S(k,a_{c})$ with respect to $k$ and $a_{c}$ to determine the maximum possible growth of a mode starting at $a_{0}$. 

From Eq. \eqref{eqn:growthfactor}, the growth factor from $a_{0}$ to $a_{c}$ is given by 
\begin{align}\label{e31}
	S(k, a_{c}) =
	\sqrt{\frac{3 A R^2}{2 m^2}}
	\frac{k}{a_{0} H_{0}}
	\int_{1}^{x_{c}} x^{-2} \left(1 - \frac{k^{2}}{k_{c\,0}^{2}} x\right)^{1/2} dx
	\,,
\end{align}
where $x_{c} = a_{c}/a_{0}$.
This does not reduce to a simple analytical expression as a function of $k$ and $a_{c}$. To obtain a conservative analytical expression for the growth factor,  we will set $x = x_{c}$ in the square root factor in the integrand. This will overestimate the suppression of the growth factor due to this  term. Then 
\begin{align}
	S(k, a_{c}) &=
	\sqrt{\frac{3 A R^2}{2 m^2}}
	\frac{k}{a_{0}H_{0}}
	\left(
		1 - \frac{k^{2}}{k_{c}(a_{c})^{2}}
	\right)^{1/2}
	\int_{1}^{x_{c}} x^{-2} dx
	\nonumber\\
	&=
	\sqrt{\frac{3 A R^2}{2 m^2}}
	\frac{k}{a_{0}H_{0}}
	\left(
		1 - \frac{k^{2}}{k_{c}(a_{c})^{2}} 
	\right)^{1/2}
	\left[ 1 - \frac{a_{0}}{a_{c}} \right]
	\,.
\end{align}
Maximizing this with respect to $k$ gives 
\begin{align}
	k = \frac{k_{c}(a_{c})}{\sqrt{2}} =  \left(\frac{a_{0}}{a_{c}}\right)^{1/2} \frac{k_{c\,0}}{\sqrt{2}} \,.
\end{align}
Therefore
\begin{align}
	S_{{\rm max}}(a_{c}) = \sqrt{\frac{3 A R^2}{2 m^2}}
	\frac{k_{c\,0}}{2 a_{0}H_{0}} 
	\left(\frac{a_{0}}{a_{c}}\right)^{1/2} 
	\left[ 1 - \frac{a_{0}}{a_{c}} \right]
	\,.
\end{align}
Maximizing with respect to $a_{c}$ gives
\begin{align}
	\frac{a_{c}}{a_{0}} = 3
	\,.
\end{align}
Therefore 
\begin{align}
	S_{{\rm max}} = \frac{1}{3 \sqrt{3}}
	\sqrt{\frac{3 A R^2}{2 m^2}}
	\frac{k_{c\,0}}{a_{0}H_{0}}
	= \frac{\sqrt{2} A R M_{P}}{m^2}
	\,.
\end{align}
Here $a_{0}$ and so $R$ should be chosen such that $R \equiv \Phi_{0}$ is as large as possible while staying consistent with a potential which is dominated by the quadratic term, where $\Phi_{0}$ is the amplitude of the $\Phi$ oscillations at $a_{0}$. We define by $r_{V}$ the ratio of the quartic to the quadratic part of the potential when $|\Phi| = \Phi_{0}$, 
\begin{align}
	r_{V} = \frac{A \Phi_{0}^{4}}{m^{2} \Phi_{0}^{2}/2} = \frac{2 A R^2}{m^2}
	\,.
\end{align}
Using this, we obtain 
\begin{align}\label{eqn:SmaxQuartic}
	S_{{\rm max}} = \left(r_{V} A \right)^{1/2} \frac{M_{P}}{m}\,.
\end{align}
Thus the condition for fragmentation, $S_{{\rm max}} > 10$, will be satisfied if 
\begin{align}
	A > \frac{100}{r_{V}} \frac{m^2}{M_{P}^2}
	\,.
\end{align}

\section{Condensate Instability in T-model $\alpha$-Attractor Inflation}
\label{sec:alp-att}

To illustrate the convenience of our general results in determining whether a given model can undergo inflaton condensate fragmentation, we next apply our results to the case of T-model $\alpha$-attractor inflation.

The potential for T-model $\alpha$-attractor inflation\footnote{E-model $\alpha$-attractor inflation has an asymmetric potential and therefore cannot be analyzed using the methods described here.} with $n = 1$ in the Einstein frame is given by  \cite{Kallosh:2013,Galante:2014ifa,Kallosh:2015} 
\begin{align}
	V &= \lambda\,\tanh^{2}\left(
	\frac{\Phi}{\sqrt{6\alpha}M_{P}}
	\right)\,.
\end{align}
In the limit $\alpha \ll N$, the spectral index $n_{s} \approx 1-2/N$ and the tensor-to-scalar ratio $r \approx 12\alpha / N^{2}$, where $N$ is the number of $e$-foldings, which are in excellent agreement with observation.

The T-model potential at small $|\Phi|$ can be expanded as
\begin{align}
	V \approx
	\frac{\lambda}{6\alpha}\left(
		\frac{\Phi}{M_{P}}
	\right)^{2}
	-\frac{\lambda}{54\alpha^{2}}\left(
		\frac{\Phi}{M_{P}}
	\right)^{4}
	\,.
\end{align}
Comparing with Eq. \eqref{eqn:QuarticPot}, we obtain
\begin{align}
	m = \sqrt{\frac{\lambda}{3\alpha M_{P}^{2}}}
	\quad {\rm and} \quad
	A = \frac{\lambda}{54\alpha^{2}M_{P}^{4}}
	\,.
\end{align}
Therefore the maximum growth factor  \eqref{eqn:SmaxQuartic} is given by
\begin{align}
	S_{{\rm max}} = \left( \frac{r_{V}}{18 \alpha} \right)^{1/2}
	\,.
\end{align}
Thus the condition for fragmentation, $S_{{\rm max}} > 10$, is satisfied if 
\begin{align}
	\alpha < \frac{r_{V}}{1800}= 5.6 \times 10^{-5} \left( \frac{r_{V}}{0.1} \right)
	\,.
\end{align}
Thus fragmentation of the condensate in T-model $\alpha$-attractor inflation will occur if $\alpha \lesssim 10^{-4}$.
Our result agrees with a numerical analysis performed in Ref. \cite{Lozanov:2016hid}, which found fragmentation for values of $\alpha$ which are consistent with this bound.\footnote{We thank K. Dimitrov-Lozanov for bringing this to our attention.}

\section{Discussion}
\label{sec:disc}

The general method and specific analytical results we have presented here are of direct significance to the general study of fragmentation of a real scalar condensate, where they are complementary to numerical studies and indeed can serve to guide such studies. 

In addition to its role in reheating, condensate fragmentation could also be important for the formation of observable gravitational waves \cite{grav1,grav2}. Our method and results can readily be applied to study fragmentation in any dominant real scalar condensate, such as that which might result from an out-of-equilibrium phase transition. They can therefore allow the parameter space of such models to be explored for the production of observable gravitational waves. We will consider further applications of our method and results in future work.


\end{document}